# A Survey on Adversarial Information Retrieval on the Web

Saad Farooq
CS Department
FAST-NU
Lahore
saad.farooq@nu.edu.pk

*Abstract*—**This survey paper discusses different forms of malicious techniques that can affect how an information retrieval model retrieves documents for a query and their remedies.**

## I. INTRODUCTION

The search engines that are available on the web are frequently used to deliver the contents to users according to their information need. Users express their information need in the form of a bag of words also called a query. The search engine then analyzes the query and retrieves the documents, images, videos, etc. that best match the query. Generally, all search engines retrieve the URLs, also simply referred to as links, of contents. Although, a search engine may retrieve thousands of links against a query, yet users are only interested in top 5 or top 10 links. The contents of links that are on the top of the retrieved list are of more importance for the given query according to the algorithms being used behind-the-scenes by search engines.

Search engines usually determine whether a document is relevant to the given search query or not by two metrics: relevance and importance. Relevance refers to the textual similarity of the page with the query. Relevance is expressed in a numerical value. Higher numerical value means more relevance. Importance refers to the global popularity of the page. Importance of a page is independent of the query, and is measured by inbound links to that page. Pages with many incoming links are considered more important.

All websites want to have their link on top of the list because it increases the traffic on their site. The process of improving the rank of a site is also called Search Engine Optimization (SEO). The SEO techniques are classified into two categories: white hat techniques and black hat techniques. White hat techniques are those techniques which are legit and are therefore approved by search engines. White hat SEO techniques actually are helpful for the web community since these techniques help the sites in structuring their site well and improving the quality of content. On the other hand, black hat SEO techniques are usually malicious and do not improve the quality of a site's content. Search engines therefore consider Black hat SEO techniques illegal, and try their best to detect and reduce the rank of or remove the links of sites altogether that practice black hat SEO techniques.

In this paper we will discuss the deceptive means (black hat techniques) that are frequently deployed by many sites to improve their rank on a search engine. The purpose of using black-hat SEO techniques may be to lure the users into visiting their page. Such pages usually are malicious and can get the user to divulge their personal information or financial details. Such pages are also referred to as spam pages.

In the end, we discuss about spam in user-generated content, including in blogs and social media.

## II. WEB SPAM

Web spamming refers to the deliberate manipulation of search engine indexes to increase the rank of a site. Web spam is a very common problem in search engines, and has existed since the advent of search engines in the 90s. It decreases the quality of search results, as it wastes the time of users. Web spam is also referred to as spamdexing (a combination of spam and indexing) when it is done for the sole purpose of boosting the rank of the spam page.

There are three main categories of Web Spam [1] [2].
1. Content based web spam
2. Link based web spam
3. Page-hiding based web spam

We will discuss these categories one by one.

### A. Content Based Web Spam

While evaluating textual similarity of a page with the query terms, search engines consider how many times and where the query terms appear on the page (location of the term on the page). There are different types of locations of a web page. Such locations are also called fields. Examples of a web page's fields are the body field, the title field, the meta tag field, the html header field, etc. Search engines give different weights to different fields.

Furthermore, the anchor text in the anchor field of links on other pages, which point to page *p*, are considered as belonging to page *p*. Anchor text is important since it usually very well describes the content of the referred page. [3]

Content based spamming exploits the above mentioned fields to increase the rank of a page for certain keywords.

### 1. Retrieval Models Affected by Content Based Spamming

Many search engines use various forms of TF-IDF metric to determine the relevance of documents for a query. TF in TF-IDF is the frequency of the term *t* in document *d*. For example, if the term ipod appears 15 times in document *d*, and document *d* has a total of 30 words, then its TF will be 15/30= 0.5. IDF in TF-IDF is related to the number of documents in the corpus in which term *t* appeared. For example, if the term ipod appears in 10 documents and there are a total of 100 documents, then its IDF will be log (10/100) = 2. The overall TF-IDF score for a document *p* and a query *q* is calculated as summation of TF-IDFs of all terms in the query.

$$\text{TFIDF}(p, q) = \sum_{t \in p \text{ and } t \in q} \text{TF}(t) \cdot \text{IDF}(t)$$

Keeping TF-IDF scores in mind, spammers can have two options. They can either make a document relevant for a large number of queries (non-zero TF-IDF score for a large number of queries) by adding many terms in the document. Alternatively, spammers can make the document relevant for a specific query by repeating the specific terms many times [1][2]. Spammers cannot influence the IDF of a term since IDF is out of their control, they can only affect the TF of a document.

**2. Content Based Web Spamming Techniques**

We will now discuss some content-based web spamming techniques.

- **Repetition of a Few Spam Terms**

Specific spam terms can be repeated in the spam document to increase its relevance for a small number of queries. The document usually has some legit content, and the spam terms are injected within the content of the document at calculated positions. Adding spam terms at calculated positions within a document is also called *keyword stuffing*.

- **Dumping a Large Number of Spam Terms**

A large number of unrelated spam terms can be dumped in a document to make it relevant for a large number of queries. Dumping is especially effective for query terms that are rare.

- **Scraping**

Scraping is a technique in which contents of many other legit pages are copied and put together into a single page. The resultant page may then show up in the search results for the terms that the original contents are relevant for.

- **Article Spinning**

Article spinning involves rewriting the original contents of different pages and putting them into a single page. This is like scraping, but the original contents are rephrased. Article spinning is usually done instead of scraping because search engines often penalize pages for duplicate contents.

**3. Where Spam Terms Can Appear in Spam Documents**

There are different positions in a web document where spam terms can appear. [1][2] [12]

- **Document Body**

Spam terms can appear in document body.

- **Document Title**

Search engines give higher weight to terms appearing in the title of documents. Spammers see this as an opportunity for spamming and put spam terms in the title of documents.

- **Meta Tag**

Search engines usually ignore the meta tags because of spamming in the meta tag. It should be noted that the terms in the meta tag do not appear on the browser, they just provide meta data about the html pages, and usually direct the browsers on how to display the contents.

Here is the example of an unspammed meta tag:

```
<head>
  <meta charset="UTF-8">
  <meta name="description" content="Web tutorials">
  <meta name="keywords" content="HTML, CSS, PHP" >
  <meta name="author" content="Sam Johnsons">
</head>
```

A spammed meta tag may look like this:

```
<meta name="description" content="cameras, Canon, Nikon, DSLR camera, DSLR photography, Digital Cameras, Digital Photography, HD Photography, photos, photography, lens, photographer, camera for vacation, photography for wedding, photography for party, photography for trip, photography for convocation, photography for annual dinner, photography for safari park, animal photos, kids photos" >
```

- **Alt Attribute**

The Alt attribute is used to provide a text description of images to search engines because search engines cannot interpret images. Spammers may add images and provide spam terms in the alt attributes in the hope that the page may get indexed for those spam terms [2].

- **Anchor Text**

Anchor text usually gives a good summary about the page which is being pointed to. Hence, search engines give high weight to anchor text. Spammers therefore create links to their spam page and add spam terms in the anchor text. This spamming technique is different from other techniques, since in this technique, the spam terms are not added to the spam page itself but to the pages that are pointing to that spam page. An unspammed anchor text may look like this:

```
<a href="https://targetlink.com">  Click Here to Further Read about WW2 <a>
```

Whereas, a spammed anchor text may look like this.

```
<a href="https://targetlink.com"> Buy cameras, Canon, Nikon, DSLR camera, Digital Cameras, Digital Photography </a>
```

- **URL**

Search engines also consider the terms being used in the URL of a document to determine its relevance. If the terms of a query exist in the URL of a document, then that document is given a higher relevance score for the query. Spammers see this as an opportunity and create long URLs

that contain the spam terms. An example of a spam URL is as follows:

photography-canon-cameras-hd-dslr-photos.net

*B. Link Spamming*

Search engines also take into consideration the link information of pages. By link information, what we mean is that for a page *p*, a search engine will check how many other important pages point to page *p*. A page is important if many other pages point to it. A page's importance also increases if another important page points to it. Suppose you have your blog page and the official website of CNN points to your blog page. In such a case, the importance of your blog page will increase since a very important site points to your site.

In link spamming, web pages are divided into following categories.

- *Inaccessible Pages* are those pages whose outgoing links cannot be changed by a spammer. Inaccessible pages are not owned by a spammer.

- *Accessible Pages/ Hijacked Pages* are those pages which are not owned by a spammer, but he can partially modify their outgoing links. For example, a spammer can add a link to his spam page in comments of a blog page. Such pages are also termed as hijacked pages.

- *Own Pages / Boosting Pages* are those pages which are owned by a spammer, and hence, he has full control over the outgoing links of own pages. The set of own pages that a spammer owns is called *spam farm* ∑. Pages in the spam farm are used to link to the page whose rank is to be increased. Such pages, which are being used to link to the target page to increase its rank, are also termed as boosting pages.

- *Target Page / Boosted Page* is the page whose ranking a spammer wants to increase. Target page is also owned by the spammer.

There are different methods for determining the importance of a page using link information. We will discuss a few methods concisely and will also discuss how these methods can be exploited by spammers to increase the importance of their pages.

**1. Hyperlink-Induced Topic Search (HITS)**

Hyperlink-Induced Topic Search, alternatively known as Hubs and Authorities, or simply HITS, is an iterative algorithm which uses link information to rank web pages.

HITS is used to rank pages on a specific topic (the importance of pages depends on the query). HITS divides pages into authority pages and hub pages. Important authority pages are those pages which are pointed to by many important hub pages. On the other hand, important hub pages are those pages which point to many important authority pages. More importance is given to pages whose hub score and authority score are higher [30].

- **HITS Algorithm Working**
  We will now give a concise explanation of how HITS algorithm works.

1. HITS algorithm first finds the most relevant pages to the search query. These relevant pages form the root set.
2. The root set is then transformed into a base set by incorporating those pages that are linked by the pages present in the root set and also incorporating those pages that link to the pages present in the root set.
3. Initially the authority score and the hub score of each page in the base set is 1.
4. The authority score of a page is found by calculating the sum of hub score of each page that points to it.
5. The hub score of a page is found by calculating the sum of authority score of each page that it points to.
6. Normalize the authority score of a page by dividing it with the square root of sum of squares of hub scores of pages that point to it. Similarly normalize the hub score of a page by dividing it with the square root of sum of squares of authority scores of pages pointed by it.
7. Repeat from step 4 until convergence or the desired number of iterations.

**2. PageRank**

Page rank algorithm is another algorithm which gives importance to incoming links of a page. The rank of a page $p_i$ is the sum of ranks of all pages $pj$ that point to $p_i$. The basic formula for page rank is as follows.

$$r(p_i) = \sum \frac{r(p_j)}{|p_j|}$$

$p_j$ is a page that points to $p_i$ and $|p_j|$ is the total number of pages that $p_j$ points to.

**3. Different Approaches to Link Spamming**

After having briefly described the working of some famous link based ranking algorithms, we are going to discuss a few link spamming techniques that are deployed by spammers.

- **Approach to Link Spamming Using Outgoing Links**
  A spammer may add a number of outgoing links to famous pages on his spam page. This way, the hub score of his page will increase. The most famous method of adding a large number of outgoing links is *directory cloning*. Directory cloning is the method of copying links from a *web directory*. A Web directory is an online list or catalog of websites. A web directory organizes links on the basis of topics and subtopics. By using outgoing links, spammers are actually trying to target search engines that deploy HITS algorithm for link analysis.

- **Approach to Link Spamming Using Incoming Links**

    A spammer wants to maximize the number of incoming links to his spam page to boost its ranking. There are different ways that can help a spammer increase the incoming links to his spam page. Some of those ways are discussed below.

    o **Honey Pot**

    Honey Pot refers to a set of pages that provide useful information to users, such as information about latest movies, or information about android programming. However, apart from such useful information, these pages also have hidden links to the spammer's target spam page. Since the honey pot appears to have useful information for users, some sites may be tricked into pointing to it, thus, indirectly increasing the ranking of the target spam page.

    o **Infiltrating a Web directory**

    Several web directories allow users to add links. If the admins/moderators of a web directory are not carefully analyzing the links posted by users, then spammers may have an opportunity to add links to their target spam page in a web directory. This technique however is slowly becoming less attractive for spammers because search engines now-a-days give less important to web directories.

    o **Link Exchange**

    Often, spammers exchange links with each other so that their spam pages point to each other, thus, mutually increasing the rank of their pages. This technique is also called *collusion attack* [11].

    o **Exploiting Expired Domains**

    When domain names expire, the pages that point to them may not be updated immediately to remove links to expired domain names. Some spammer may buy an expired domain and add their spam pages to it. Thus, the pages that are pointing to an expired domain name will unintentionally be pointing to a spam page. This technique is particularly useful for spammers because certain search engines consider the domain age for determining the rank. Older the age of a domain name, better the rank. The reason that domain age is considered by search engines while ranking, is that a spam page cannot stay up for a long time. Users eventually find out spam pages. So, the expiry of an old domain name is in the best interest of spammers. Furthermore, an expired domain name must previously have some user traffic. So, a spammer can get advantage of this traffic as well [4].

    o **Posting Links to Blogs, Wikis, and Message Boards**

    Spammers can post links to their spam pages in blogs, wikis and message boards, etc. as part of their apparently innocent post. If the blog, wiki, or message board is un-moderated, then it may end up linking to spam pages. If a blog site allows embedding html within the blog post, then spammers deploy content hiding techniques for adding links. As a result, even if the blog is moderated, it sometimes becomes difficult for moderators/editors to detect spamming because of hiding techniques. Content hiding techniques for spamming have been discussed later in this paper.

    o **Link Bombs**

    The fact that search engines take into consideration the links and their anchor text for ranking pages has been exploited many times. A phenomenon known as link bombing is the collaborative effort to get a search engine to rank the target web page highly in the result for a certain search query. If link bombing is being done for Google search engine, then it is called a *Google bomb* [11]. There can be many motivations behind this, such as humor, ego, ideology, or revenge. The link bomb organizers attempt to convince many web sites to use particular anchor text in a link to a particular target page, so that the target page is ranked highly by search engines for the terms that appear in the anchor text.

    Link bombing is often done for the purpose of online protest [12] [13].

    Link bombing has earned fame a number of times in the popular press [12]. The first link bombing attempt and the source of the term *Google bomb* was Adam Mathes' 2001 Blog post which encouraged the creation of a Google bomb for a friend's blog using phrase "talentless hack"[13]. By 2006, this technique was being used for political purposes.

    A screen shot of Google bombing in 2006 is shown below. It can be seen that for query "miserable failure", Google is returning web pages associated with George W. Bush and Michael Moore. It is one of the most famous examples of Google bombing. (Source: Wikipedia)

    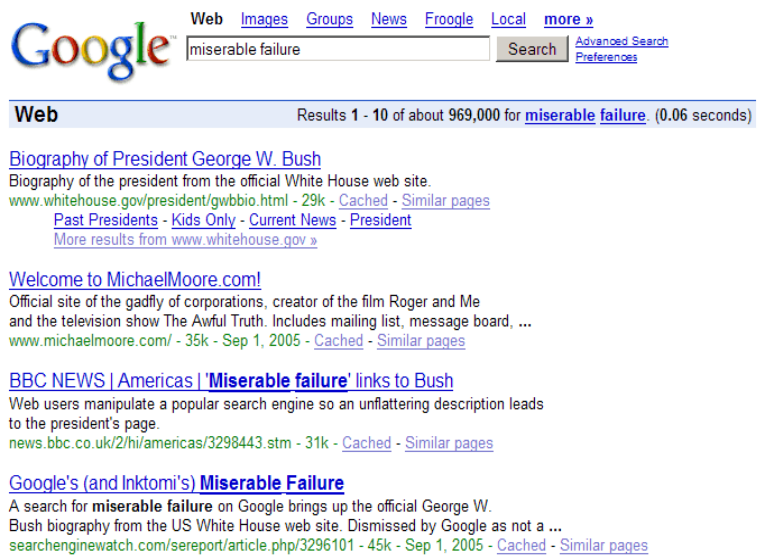

    Google bombing has become a lot more difficult today. However, Google bombing still pops up sometimes. A recent example of Google bombing is that the search for the term "idiot" in Google images showed pictures of Donald Trump [14].

    o **Link Farm**

    A link farm is a collection of pages in which each page points to the target page in order to artificially increase its rank. The page for which the spammer wants to increase the rank is called *boosted page*, and the pages which are being

used to link to the boosted page to increase its rank are called *boosting page* or *hijacked page*. Boosting pages are those pages which are under the full control of the spammer. Hijacked pages are those pages which are not under the control of the spammer, but he can partially change that page, such as the page may be a blog page and allows posting comments. In such a case, the spammer can post the links to the boosted page in comments [11].

- **Referrer Spam**

The *HTTP referer* (a misspelling of referrer) is an *HTTP* header field that identifies the address of the webpage that linked to the resource being requested. Whenever a user clicks a hyperlink on some web page on a browser, the browser sends an *HTTP* request to the server of the page, whose URL exists on the hyperlink, to fetch that page. In the *HTTP Referer* field, the address of the last page the user was on (the one where the user clicked the hyperlink) is included [25].

Web masters usually log referrers to identify what sites are referring to their pages. The log is used for promotional or statistical purposes.

Referrer spam (also known as referral spam or log spam) is a spamdexing technique (spamming meant for search engines). This technique makes use of the *HTTP Referer* field. In this technique the spammer makes web page requests using a fake referrer URL of the site that spammer wishes to advertise. If the web site, whose page is being requested, publishes its access logs, including referrer statistics, then it will unintentionally link back to the spammer's site. If the access log of a web page is public, then the web crawler of a search engine will crawl the access log as well, thus improving the spammer's ranking [26].

*C. Hiding Techniques for Web Spamming*

Hiding techniques refer to the methods in which spammers conceal their spam terms or spam links [6]. Hiding techniques are deployed along with the previous two techniques (content-based spamming and link-based spamming). There are different hiding techniques which are discussed below.

**1. Content Hiding**

In content hiding techniques, spam terms are present on the page but are not visible to the user on the browser. Invisibility of spam terms can be achieved if the color of spam terms is the same as the background color of the page.

```
<body background="black" >
   <font color="black"> spam terms </font>
</body>
```

Similarly, spam links can be hidden by using an image instead of anchor text. Spammers use tiny 1x1-pixel anchor images that are either transparent or background-colored.

```
<a href="target.html">  </a>
```

**2. Cloaking**

In this technique, spammers detect whether the current client visiting their site is a web crawler. If it is a web crawler, they return a different page to the crawler as compared to the page returned to normal users. Using this method, spammer can show their ultimately intended page to the user without any traces of spam. The page returned for indexing to the web crawler contains spam term.

A web crawler can be identified in two ways. One way is to keep track of the IPs used by popular search engines. Whenever a new client connects with the spammer's server, he can match the IP and identify whether a client is a normal web browser or a web crawler. The other way is to use *HTTP* header. In *HTTP* header, there is a field called *user-agent*. The *user-agent* field specifies the type of the browser being used to request the page. A typical *HTTP* header is shown below:

GET /somedir/page.html HTTP/1.1

Host: www.nu.edu.pk

User-agent: Mozilla/5.0

If the client visiting the site is a web crawler, then it will not be using a web browser. Web crawlers usually fill the *user-agent* field with some different value. So, using this field, spammers can detect whether the client visiting their site is a web crawler or not. Hence, they can return a different page, containing spam terms, to the web crawler. If the page returned to the web crawler gets indexed, then, whenever users search for terms which were in the spammed page, the search engine will also return the page of the spammer. However, when the user requests for this page, the page returned to the user will be a different page without any traces of spam. The reason for using cloaking is that if the same page, which is returned to normal users, were also returned to the web crawler for indexing, it would not have enough ranking power of its own. So, another page containing spam terms is returned to web crawlers which may have some ranking power.

It, however, should be noted that the method of identifying a web crawler is not only used for spam hiding, but it is also used for legit purposes. For example, if a web page has a lot of graphical elements, then there is no benefit of returning the graphical elements to the web crawler because search engines only index the text of the document. Hence, by identifying the web crawlers, web masters can send the page which contains only the text and does not contain any graphical elements, such as images, videos, advertisement, etc.

**3. URL Redirection**

A convenient and cheap alternative to cloaking is URL redirection. In this method, a browser is redirected to another URL as soon as the page containing spam terms is loaded. This way, both the crawlers and web browsers are returned the same page containing spam terms. But users will not be able to see the page on the browser because of immediate redirection. Spam Pages with redirection act as intermediate doorways for the ultimate target page, which spammers attempt to display to the users who have landed on their site through a search engine.

URL redirection can be conveniently achieved in a number of ways. First way is to use refresh meta tag in the header of an html page. The refresh time is set to 0, and the refresh URL is set to the URL of the target page. Web browsers are redirected to the target page as soon as the spam page gets loaded in the browser.

```
<meta http-equiv= "refresh"  content="0;
       url=target_page.html" >
```

URL redirection using refresh meta tag is very easy to implement. However, search engines can easily detect such URL redirection attempts while parsing the page. Pages with immediate URL redirection are often penalized by search engines. Another way to perform redirection is to use scripts. Scripts are not executed by web crawlers. Hence, URL redirection attempt can be concealed. An example of URL redirection using a script is as follows:

```
<script language="javascript">
<!-- location.replace("target_page.html") -->
</script>
```

### III. WEB SPAM DETECTION

We now present techniques for web spam detection. Just like we divided web spamming into different categories, we will also divide web spam detection techniques into different categories.

At first, we will discuss some features (content-based, URL-based, and link-based) which could be used in classifying pages as spam or non-spam.

#### A. Feature Selection For Web Spam Detection

Different features, which could be used in classifying pages as spam or non-spam, are presented in this section.

- **Content-based Feature Selection for Spam Detection**

We will now discuss different content-based features that could be selected to detect web spam.

**1. Number of Words in a Page**

While discussing content-based spamming, we discussed that a few terms can be repeated in the document, or a large number of terms can be dumped into the document. This way, the length of the document increases. Hence, the document length may be used as a heuristic to determine whether the document is a spam document or not. We now discuss the experimental results of [5].

The authors of [5] plotted a graph with respect to the number of words in each page of their dataset. The graph is shown in figure 1 below.

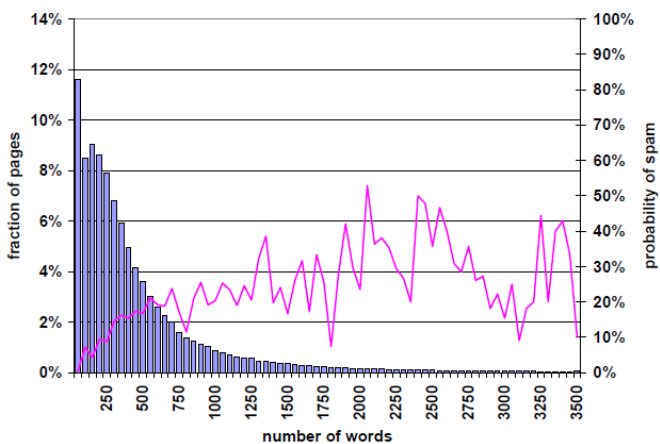

Figure 1: Spam Prevalence Relative to Word Count [5]

From figure 1, it can be observed that there does exist some correlation between the number of words on the page and spam prevalence. However, percentage of spam pages in each range is below 50%. So, word count alone cannot be used to identify a page as a spam page. We need to incorporate other heuristics as well. Furthermore, we need to give less importance to number of words on the page for the detection of spam pages; otherwise there will be many false positives.

**2. Number of Words in the Title of a Page**

Search engines give higher weight to terms occurring in the title of pages. As a result, spammers also add spam terms in the title of their pages. The author of [5] carried out the investigation whether the number of words in the title of a page is a good indicator of a spam page or not. They have plotted a graph of their results as shown in figure 2.

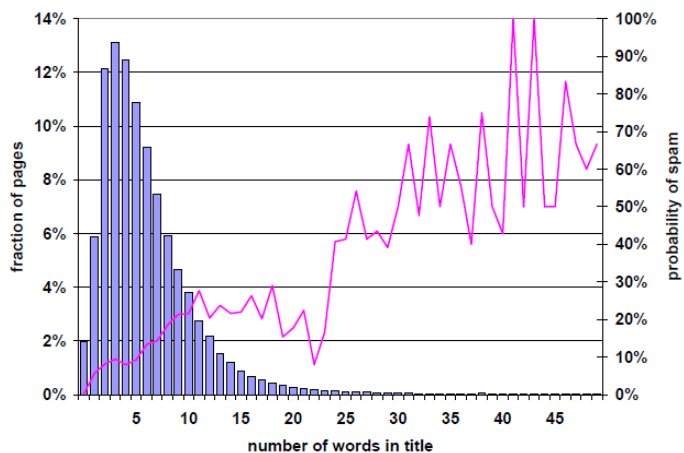

Figure 2: Prevalence of Spam Relative to Word Count in Title of Pages [5]

From figure 2, it can be seen that most of pages in the dataset used by the authors of [5] contain less than 10 words in their titles, and the probability of spam increases when the number of words in the title of a page increases.

So, we can conclude that the number of words in the title of a page is a good indicator of whether a page is a spam page or not. As a matter of fact, the number of words in the title of a page is a better heuristic than the number of words on the whole page for detecting spam.

### 3. Fraction of Anchor Text

As discussed earlier, anchor text usually gives a good summary about the page that is being pointed to. For example, if page *B* is pointing to Page *A*, and the anchor text is "World war 2", then page *A* is probably written about World War 2.

Spammers exploit this opportunity and create spam pages that exist solely for providing anchor text to target pages. The authors of [5] investigated whether the amount of anchor text on a page is a good heuristic in identifying spam. For every web page in their data set they calculated the fraction of all terms (excluding markup text) present in anchor text. The results are shown in figure 3. From the figure, it can be seen that probability of spamming of a page increases with an increase in the fraction of anchor text present on the page. However, the probability is hardly more than 50% for the fraction amount (of anchor text) where probability of spamming is maximum. So, just like the number of words on a page, using anchor text alone as a heuristic to identify spam may cause false positives. However, when this heuristic is combined with surrounding text and URL terms, it works better.

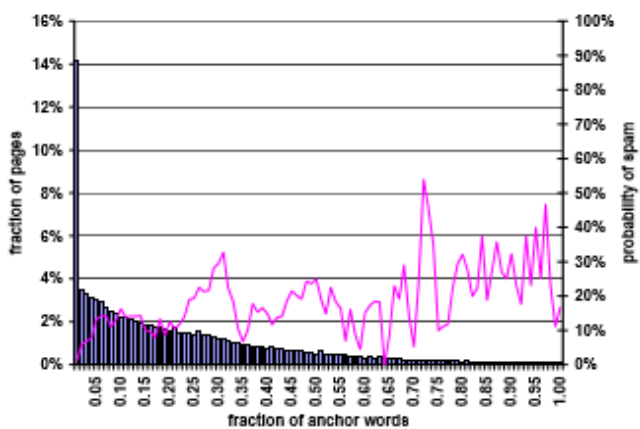

Figure 3: Prevalence of Spam Relative to Fraction of Anchor Text [5]

### 4. Fraction of Visible Content

The visible content of a page is the content that is visible to the user on a web browser. Whereas, the invisible content of the page is the content that is not visible to users on the browser, such as html tags, html tag attributes, executable scripts, CSS blocks, etc. The author of [5] investigated whether the fraction of visible content is a good heuristic in identifying spam. For every page in their data set, they calculated the fraction of visible content, by dividing the total amount of visible content with the total size of page. Figure 4 depicts the distribution of fractions of visible content.

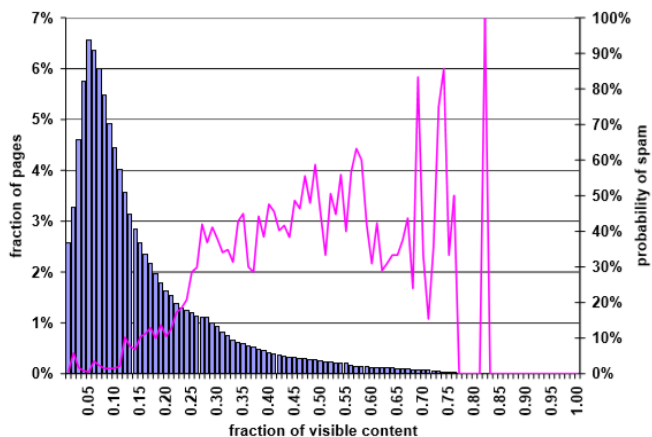

Figure 4: Prevalence of Spam Relative to Fraction of Visible Content [5]

The line graph, which shows the probability of spam against the corresponding fraction of visible content, rises as the fraction of visible content increases. The line graph ends at 82% because in the dataset used by the authors, there is no page which has more than 82% of visible content. It can be seen from the figure that an increase in the fraction of visible content of a page increases the probability of spamming. So, we can conclude that spam pages usually contain less mark-up, scripting, or styling language. It is obvious because many spam pages are created so that they can be indexed by search engines. Spam pages are not meant to be useful for humans. As a result, little effort is made to create an interactive and graphically pleasing page [8].

### 5. Duplicate Content

Spammers often copy contents from famous pages in the hope that their spam page will also be given higher rank for the queries that the copied contents are relevant for. So, if the content of a page is a duplicate of some other page(s), then there is a good chance of spamming.

### 6. Content Compressibility

Spammers often repeat the terms on their spam page. Repetition increases the compressibility of a page. Hence, compressibility is a good measure for the detection of spam [7].

### 7. Number of Advertisements

Spammers also create pages for the purpose of generating profits, so they often put too many advertisements on their page. Presence of unusual number of advertisements on a page may allude to spamming [8][9][7].

### 8. Too Many Call-to-Action Phrases

If a page contains high number of call-to-action phrases, such as limited offer, hurry now, only x items left in stock, offer closes on 'date', only for first 50 customers, etc., then it alludes to monetization of page, promotion of scam, malware, or clickbait, etc. [8]

9. **Image Count**
High quality pages normally have adequate number of images. Spam pages are low quality pages created using tools and therefore contain no or very few images. So, image count can be combined along with other heuristics to identify spam [9].

10. **Low Count of Stop Words**
Pages created with keyword stuffing usually have lower number of stop words [8][9].

- **URL-based Feature Selection for Spam Detection**
Spammers create URLs in bulk using automated tools. Such URLs are often used solely for the purpose increasing rank of their target page(s). Sometimes, such URLs are not meant to be visited by users. They are created for the sake of increasing the rank of a spammer's target pages. For example, such URLs may be part of link farm, so the page of a spammer's URL may contain links to the target page to artificially increase the rank of the target page. Hence, there can be many abnormalities with spam URLs. However, it should be noted that spammers may also create URLs, which are meant to be visited by users, using automated tools. We will now present URL-based features that could be selected to detect spam.

1. **SSL Certificate**
SSL stands for Secure Socket Layer. SSL is used to establish a secure connection between the browser and the server. An SSL certificate ensures that the website is trusted. SSL certificates require extra cost. So, spammers do not use SSL certificates for their spam site [8]. It should be noted, that many legit sites also do not use SSL certificates, so using the absence of SSL certificates alone as a heuristic can cause false positives. The presence/absence of SSL can however be combined with other heuristics to identify spam sites.

2. **URL Length**
Spam sites usually have longer URL length due to keyword stuffing in the URL, so larger URLs can be a signal of spam [8][10]. Search engines also prefer short URLs [10].

3. **Multiple Sub-domains of URL**
Spammers often create multiple sub-domains on a single domain to create multiple websites. By doing so, spammers save on the cost of purchasing multiple domains. So, spam pages have high probability of being hosted on sub-domains [8]. For example, blog.johnson.me has three levels of domain, .me is the top level, .johnson is the second level, and blog is the third level.

4. **Authoritative Top-Level Domain**
There are certain authoritative top-level domains that can only be registered by a legally recognized authority. The examples of authoritative top-level domains are .edu, .gov, etc. Authoritative top-level domains cost more. So, it is unlikely that spammers will use authoritative top-level domains to host their spam sites [8][10].

5. **IP Address Instead of Domain Name**
Spammers often host their spam pages on bare IPs instead of purchasing domain name to save cost [8]. An example of a page address hosted on a bare ip is: 128.154.125.123/home.html.

6. **Name of a Popular Company in Sub-domain**
Spam URLs often have the name of a popular company so that a naïve user may think that the URL belongs to that company, such as apps.facebook.login.co. Such URLs are also often used for phishing attacks.

7. **Many Digits or Special Symbols in URL**
URLs are kept user-friendly. So, legit sites often refrain from using many digits or special symbols in their URLs. However, spam URLs are often generated by automated tools which may produce low quality URLs. So, if there are many digits or special symbols in a URL, then the URL may belong to a spam site.

8. **More than 2 Consecutive Same Characters in URL**
As discussed previously, Spammers create URLs in bulk using automated tools. Tools may generate URLs that contain the same character consecutively more than 2 times, such as buy-carsss.me. So, having the same character more than 2 times consecutively is often a signal of spam.

- **Link-based Feature Selection for Spam Detection**
We will now present link-based features that could be selected to detect spam.

1. **Number of Internal Links**
A user-friendly website has a good number of internal links to make navigation easy for users [8]. Spam pages often lack adequate number of internal links.

2. **Self Referencing Links**
Sometimes, spammers put too many self referencing links. Each link has different keyword in its anchor text to boost its rank for different keywords. So, the presence of many self referencing links with different anchor text may be a signal of spam.

*B. Web Spam Detection Techniques*
We will now discuss some spam detection techniques in web pages.

1. **Content-based Spam Detection Techniques**
Content-based spam detection uses content-based features to detect spam. We will now discuss some content-based spam detection techniques in this section.

- **Spam Detection using N-gram Language Model**
Apart from abnormalities in the title, anchor text, etc. of spam pages, the authors of [5] discovered that spam pages have an abnormal language model, including the fact that spam pages have a greater number of popular terms than normal pages. They built a 3-gram of language model for their dataset and discovered that spam pages have abnormally high or abnormally low likelihood of a query.

This is because distribution of n-grams of spam pages is substantially different from background distribution.

An augmented representation of textual content of pages can be used to improve the accuracy of classifying pages as spam or non-spam pages. The authors of [15] annotated the documents with part-of-speech (POS) to find out the morphological class of each word, e.g. verb, noun adjective, etc. The authors discovered that the sequence such as *<noun, verb, noun>* is more likely to occur than *<verb, verb, verb>* in non-spam English pages [15].

- **Spam Detection Using Generative Models**

An alternative technique for finding spam pages is to use generative models, such as Latent Dirichlet Allocation (LDA). The philosophy of LDA is that when writing a document given a language model, the author first picks a topic according to a distribution over topics, and then picks a term according to a topic-dependent distribution over words. The authors of [16] used a multi-corpus LDA to find whether a document is more likely to have been generated from a spam-model or a non-spam model. Both spam and non-spam models are created by a training set. In the training set, the documents are already labeled as spam or non-spam.

The results of classification using Linked LDA can be seen in the table below. *k* is the number of topics, and *p* is a normalization factor. For *k*=30, and *p*=4, the results were the best. Linked LDA model extends the LDA model to incorporate the effect of a hyperlink between two documents on the topic and term distributions.

|  | P=1 | P=4 | P=10 |
|---|---|---|---|
| K=30 | 0.768 | 0.784 | 0.783 |
| K=90 | 0.764 | 0.777 | 0.773 |

Table 1: Classification Accuracy in AUC for Linked LDA with various parameters, classified by BayesNet [16]

| Features | AUC |
|---|---|
| LDA with BayesNet | 0.766 |
| Tf.idf with SVM | 0.795 |
| Public (link) with C4.5 | 0.724 |
| Public (content) with C4.5 | 0.782 |

Table 2: Classification Accuracy in terms of AUC for baseline methods [16]

| Features | AUC |
|---|---|
| Tf.idf & LDA | 0.827 |
| TF.idf & Linked LDA | 0.831 |
| Public & LDA | 0.820 |
| Public & Linked LDA | 0.829 |
| Public & TF.idf | 0.827 |
| Public & Tf.idf & LDA | 0.845 |
| Public & Tf.idf & Linked LDA | 0.854 |
| Public & Tf.idf & LDA & Linked LDA | 0.854 |

Table 3: Classification Accuracy Measured in Terms of AUC by Combining Different Classifiers [16]

- **Classifying Pairs of Documents for Spam Detection**

Some researchers have worked on detecting nepotistic links on a page using some or all of the content of the source and target pages. The supposition behind this approach is that in a non-spam link, the content of the source page and the target page should be similar [12].

The authors of [11] and [17] measure the Kullback-Liebler divergence of the unigram language model of both the source page and the target page. If divergence is high, then the link from the source page to the target page is considered as nepotistic. The authors of [17] state that computing divergence of all pairs of documents connected by a link may be computationally expensive. So, they have suggested comparing only the anchor text in the source document with the target document.

The same method can be applied to detect spam comments in blogs by analyzing the divergence between language model of the comment and the language model of the blog on which the comment has been posted [18].

## 2. Link-based Spam Detection Techniques

There are many methods for detecting link spam. We will discuss some of them in this section.

- **TrustRank**

The authors of [27] have proposed an algorithm "TrustRank" for the detection of link spam. The basic assumption in TrustRank is that good pages always point to good pages; they seldom point to bad pages (pages containing spam).

In TrustRank, some trusted pages are selected as seed set and trust scores are assigned to them; the remaining pages are initially assigned a trust score of 0. In selecting pages for seed set, more preference is given to pages with many outlinks, so that trust can propagate quickly to many pages. For the selection of seed set, the authors computed the inverse PageRank of each page. Inverse PageRank is computed by reversing the in-links and out-links in the webgraph. In other words, the original PageRank algorithm is run on the transpose of the web graph matrix. A higher inverse PageRank value of a page indicates that more pages can be reached from this page in fewer hops. So, pages with higher inverse PageRank scores are selected for seed set. Such pages, however, need to be examined by human experts for spam.

TrustRank algorithm works by propagating trust scores of pages from seed set to all other reachable pages on the Web. In the end, the pages with high trust scores are considered as good pages, and the pages with low trust scores are considered as spam pages. TrustRank computes trust score in the same way the original PageRank algorithm computes the page rank score. Just like PageRank algorithm, TrustRank also includes the concepts of random surfer model with teleportation. In TrustRank, by teleporting, the random surfer always jumps to one of the pages in the seed set. [27] .

- **Anti-TrustRank**

Anti-TrustRank is based on the same supposition as TrustRank, i.e., a good page always points to good pages and seldom points to a bad page. Anti-TrustRank, however, does not find the trust scores; it finds the anti-trust or badness scores of pages. The higher the badness score of a web page, the higher the probability that the page is a spam page.

The algorithm works by first finding a seed set containing spam pages. Anti-TrustRank is then run on the transpose of web graph. Anti-TrustRank algorithm also includes the concepts of random surfer model with teleportation. Using teleportation, the random surfer will always jump to one of the pages in the seed set. In the end, pages with higher anti-Trust scores are considered to be spam pages [28].

### 3. Combining Content-based, Link-based, and URL-based Features to Detect Spam

For the detection of spam, one could use a number of content-based, link-based, and URL-based features to detect spam using supervised machine learning approach. We have previously described a number of content-based, link-based and URL-based features that could be used in detecting spam. All those features will be fed to a classifier. The classifier will then be trained using training data.

The authors of [8] used 32 features in total to detect spam using classification method. They used a classifier for the detection of spam using 32 features of a web page. The classifier was based on Resilient Back-propagation Learning algorithm based multilayer neural network. The size of dataset was 370 pages. In their dataset, 30% of pages were spam pages. They randomly selected 300 pages from the set for training. The remaining 70 pages were used for testing the classifier accuracy. The results of evaluation carried out by authors are given in table 4.

| Features | Efficiency | Precision | Accuracy |
|---|---|---|---|
| URL | 0.7153 | 0.8460 | 0.7433 |
| Content | 0.8154 | 0.8489 | 0.8267 |
| Link | 0.7731 | 0.8329 | 0.7900 |
| URL+ Content | 0.9092 | 0.9403 | 0.9155 |
| URL+ Link | 0.9028 | 0.9500 | 0.9111 |
| Content+ Link | 0.8204 | 0.8540 | 0.8308 |
| URL+ Content + Link | 0.9168 | 0.9357 | 0.9216 |

Table 4: Results of Evaluation of Classifier for the Detection of Spam using Different Page Features used by authors of [8]

From table 4, it can be seen that when all categories of features (URL, Content, and Link) are deployed, the accuracy of classifier in detecting spam become maximum.

### 4. User Behavior-based Spam Detection

The authors of [32] have proposed another approach which uses the user behavior on a spam site. They have proposed three user behavior-based features to detect spam pages.

Authors have proposed that since a spam page's content is not useful, a web spam page therefore receives most of its user visits from search engines instead of from clicks on a hyperlink on non-spam pages or from bookmark lists. They defined the *Search Engine Oriented Visit Rate (SEOV)* of a page $p$ as follows:

$$SEOV(p) = \frac{\#(Search\ engine\ oriented\ visits\ of\ p)}{\#(Visits\ of\ p)}$$

According to authors, a spam page will only be visited through some search engine. A useless page will seldom be directly visited by users. *Search Engine Oriented Visit Rate* measures the total fraction of visits that happened through a search engine.

Furthermore, they have opined that whenever users visit a spam page, they don't generally follow hyperlinks on that page. In other words, their navigation on spam sites end as soon as they notice that the site is a spam site. So, the authors have proposed another feature which calculates the ratio between the number of clicks on a hyperlink on a page $p$ while visiting it and the total number of visits of $p$. They called this feature *Start Point Visiting rate (SP)*.

$$SP(p) = \frac{\#(user\ clicks\ a\ hyperlink\ on\ p\ while\ visiting\ p)}{\#(Visits\ of\ p)}$$

The third feature that the authors have proposed is based on the opinion that generally web sites want to keep users navigating on their site. For this purpose, websites provide a number of internal hyperlinks. If a site is a spam site, then users will not follow internal hyperlinks; just like they will not follow external hyperlinks. Authors have called this third feature as *Short-Term Navigation Rate (SN)*. The short-term navigation rate measures how many pages of a site $s$ will be visited once the user visits $s$.

$$SN(s) = \frac{\#(Sessions\ in\ which\ users\ visit\ less\ than\ N\ pages\ in\ s)}{\#(Sessions\ in\ which\ users\ visit\ s)}$$

In order to evaluate the accuracy of the above mentioned three features, the authors collected Web access log from July 1, 2007 to August 26, 2007 of sohu.com using browser toolbars. The access log has information of more than 2.74 billion user clicks on 800 million Web pages and 22.1 million user sessions during 57 days. Some statistics observed by authors in their dataset are as follows:

1. Over 80% non-spam pages get less than 10% of visits through search engines.
2. Around 50% of spam pages get 50% of visits through search engines.
3. 63% of non-spam pages get more than 30% user clicks on any hyperlink on them.
4. 48% of spam pages get less than 5% user clicks on any hyperlink on them.

After observing the statistics, the authors have proposed that *Search Engine Oriented Visit Rate* alone can be used to detect spam pages. It is due to the fact that a spam page's main target is the search engine. In some cases, a search engine is the only way a spam page can be visited. The authors used naïve Bayes classifier which used the above mentioned three features to detect spam pages. The results of classifier are given in figure 4 below.

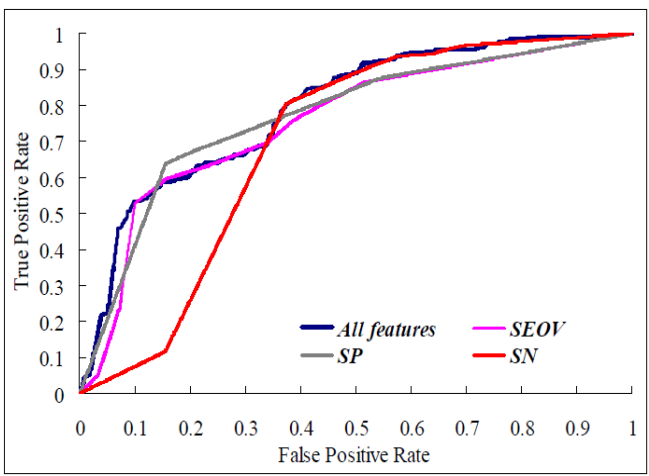

Figure 4: ROC curves on data set using Naïve Bayesian which used three features: *SEOV, SP* and *SN* features. [32]

## IV. SPAM IN USER GENERATED CONTENT

User-generated content in social media, as opposed to professionally generated content from traditional media, has been one of the main driving forces behind the growth of the Web since the early 2000s. As time goes on, more and more users participate in content creation, rather than just consumption [12]. People are becoming producers as well as consumers: "prosumers", a term which was coined by Alvin Toffler in 1980 [12].

Some of the popular user-generated content domains include blogs, photo and video sharing sites, and social networking sites, such as twitter, facebook, etc. Such platforms are used by many people for constructive purposes, but also abused by a few malicious users for deceptive or fraudulent purposes.

### A. User-Generated Content Platforms

There are three kinds of platforms for user-generated contents, which are exploited by spammers and other malicious users.

o **Free Hosting Sites**
Blogs are often hosted on sites that offer free blog hosting and creation tools. Spammers exploit such free hosting sites to create *splogs,* meaning spam blogs.

o **Publicly Writable Pages**
Publicly writable pages are pages such as opinion forums and comment forums, user review sites, and collaborative editing tools known as wikis. Wikis are websites which provide information to users about different topics. All information is publicly editable, so it gives great opportunities to spammers for adding links to their spam sites. For example, a spammer could add some information on a topic in a wiki, and add a link to his target page as reference. References usually provide the source of information on wikis. Wikipedia is a good example of wikis.

o **Social Media Sites**
Social media sites include sites where a user can upload images, videos, answers, etc., and interact with the content through votes, comments, tags, etc. Furthermore, social media sites also allow people to interact with each other through social networking features [12]. Social Media sites are most vulnerable to spam, phishing attacks, hate speech, etc.

### B. Types of User-Generated Content Spam

- **Splogs**
Spam blogs are one of the prevalent types of spam pages. Spam blogs are also called *splogs*. Some *splogs* are simply spam sites hosted on free hosting sites, and just like other spam pages; their goal is to unfairly boost the raking of some set of pages [18]. Some splogs sites also attempt to trick users into clicking on ads, or masquerade as independent opinion sources about a product or service. Splogs are often written using content spinning software. Furthermore, Splogs may contain a lot of advertisements for the generation of revenue, and an excessive amount of links to target pages to increase the rank of those pages.

- **Comment Spam**
Spammers often post links to their spam pages in comments on publicly writable page, such as blogs, wikis, forums, online guest books, etc. Spammers usually use a computer program which automatically posts comments many times. Such a program is also called a *spambot*. A comment spam may include unsolicited advertisement, links to some external page, etc. Links to external page are usually added by spammers in comments to boost the rank of that page.

- **Review Spam**
Product or service reviews of other users in general help a user decide whether a product is good or not. However, it can only help the user when the reviews are honest and are made by the actual customers. Sometimes, users may post fake reviews. The motive behind a fake review is to either convince other users into buying the product or deter them from buying the product. A user may post a negative review

just to malign the product or its manufacturer. Review spam can be divided into different categories [19]

o *Untruthful Reviews/False Reviews* may mislead users by giving undeserving positive review to a product or by giving maliciously negative review.

o *Non-reviews* do not contain any review of the product; they either contain advertisements or some other irrelevant text, such as a question, answer, or some random text.

o *Brand Reviews* are reviews that are not about the product but about the brand or the manufacturer of the product.

- **Spam in Social Media**

The exponential growth of social networking sites, such as facebook, twitter, etc., has dramatically increased the opportunities for users to find and socialize with more people. Social media platforms are most prone to spam. Comment spam and review spam also occur in social media. Apart from spam, hate speech is also very prevalent in social media. Some other types of spam that occur in social media are as follows.

o **Hashtag Hijacking**

A hashtag is a type of metadata tag which is used on social sites, such as twitter, facebook, etc. A hashtag allows users to apply dynamic, user-generated tags which make it possible for other users to easily find messages or posts on a specific theme or topic. Users put hashtags in their posts on social media by using # symbol. Searching for a specific hashtag will return all posts in which that specific hashtag has been used.

Hashtag hijacking refers to the use of a hashtag for the post whose theme or topic is contrary to what the hashtag is actually meant for. Spammers usually do hashtag hijacking for the purpose of advertisements or gaining attention.

On some social sites, such as twitter, users can see which hashtags are trending. Spammers usually take advantage of trending hashtags, and put those trending hashtags in their spam tweets [23]. So, when a user searches for tweets that are using a particular hashtag, apart from seeing legitimate tweets, he will also see those spam tweets.

o **Voting Spam**

On question-and-answer websites, such as Quora, Yahoo Answers!, etc. users ask questions and give answers to questions asked by other users. Users can also give vote to answers. Voting helps other users determine whether the answer is useful or not. However, malicious users can exploit voting system and may give undeserving positive vote to an answer to unfairly boost the importance of that answer. They may also give maliciously negative vote to an answer to degrade that answer.

V. SPAM DETECTION IN USER-GENERATED CONTENT

We will now discuss different spam detection and prevention techniques for user-generated content.

A. CAPTCHA

CAPTCHA stands for Completely Automated Public Turing Test to Tell Computers and Humans Apart. A CAPTCHA contains some text or images with noise. Human users have to identify the text or image contained in the CAPTCHA. Computers cannot detect the text or image in the CAPTCHA due to noise. This technique can slow down automatic registration and automatic posting of content [20].

CAPTCHAs do not help to detect spam, but they can help to slow down spamming process since many spammers use automated tools, such as *spambots*, to post spam comments or create spam blog pages. Using automated tools, a spammer can post multiple comments or create multiple spam blogs in a short period of time. A CAPTCHA helps to alleviate this problem.

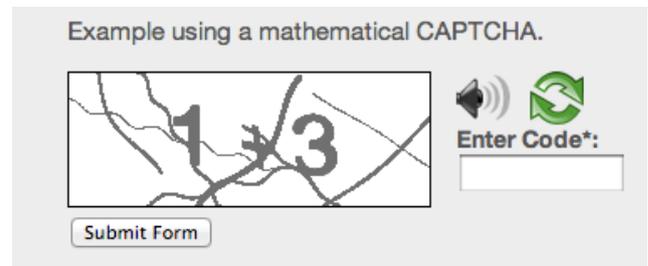

Figure 5: A CAPTCHA which asks the user to enter the result of a small mathematical expression (source: examples.com)

B. Comment Spam Detection

One of the simplest methods to detect a spam comment is to have users flag a comment as spam. If many users have flagged a comment, then it is probably a spam comment. In this section, we will discuss some techniques to automatically detect spam comments.

The authors of [18] have proposed a content-based approach to detect a spam comment. According to authors, in case of e-mail spam detection, each email should be analyzed independently. However, in case of comments there is a context, i.e., the page and the site where the comment has been posted. They have proposed a method in which the language model for the page on which the comment has been posted and the language model of the comment itself are determined. Then, the distance between the language model of the comment and the language model of the page is computed. This distance is measured using Kullback-Leibler divergence. In the end, a threshold value in the distance is used to determine whether a comment is a spam comment or not.

If contents of the page and the comment are short, then the models of both the page and the comment can be augmented by including the contents of linked pages. The language model of the page can be augmented by adding the contents of those pages that are linked by the page. Similarly, the language model of the comment can be augmented by adding the contents of those pages which are linked by the comment. This technique is especially useful when the comment is being used for the purpose of boosting the rank of the linked page(s), and in general, most of the spam comments have links to external pages. Such comments have been referred to as *link-spam comments* in [18].

Furthermore, a spammer may write the comment in such a way that it appears to be relevant to the post. In such a case, including the language model of the page linked by the comment is surely going to help because a link to an irrelevant page in a comment can increase the distance between the comment and the page on which the comment has been posted. This way, the probability that the comment is a spam comment increases [12].

The results of experiment carried out by authors of [18] for identifying spam comment using Kullback-Leibler divergence are given in table 5. For carrying out the experiment, the authors collected 50 random blog posts. The total comments posted on those blog posts were 1024. All of the collected blog posts had a mix of spam and non-spam comments. The authors removed duplicate as well as near-duplicate comments. The number of comments per post ranged between 3 and 96. They manually classified the comments as legitimate comments (32%) and link-spam comments (68%). The authors used only the language models of the post and the comments. They did not take into consideration the models of pages linked by the comments and by the page on which the comment had been posted due to time constraints.

| Method | Threshold Multiplier | Correct | False Negatives | False Positives |
|---|---|---|---|---|
| Baseline (avg. 100 runs) | N/A | 581 (57%) | 223 (21.5%) | 220 (21.5%) |
| KL-divergence | 0.75 | 840 (82%) | 65 (6.5%) | 119 (11.5%) |
| KL-divergence | 0.90 | 834 (81.5%) | 69 (6.5%) | 121 (12%) |
| KL-divergence | 1.00 | 823 (80.5%) | 88 (8.5%) | 113 (11%) |
| KL-divergence | 1.10 | 850 (83%) | 87 (8.5%) | 87 (8.5%) |
| KL-divergence | 1.25 | 835 (81.5%) | 111 (11%) | 78 (7.5%) |

Table 5: Result of Experiment carried out by authors of [18] for spam comment detection using KL divergence

Increasing the threshold multiplier increases the threshold value of divergence allowed between the comment and the page on which the comment has been made. On the other hand, decreasing the threshold multiplier decreases the threshold value of divergence allowed between the comment and the page. Increasing the threshold will result in more false negatives, whereas decreasing the threshold will result in more false positives.

*C. Review Spam Detection*

For the detection of non-reviews and brand reviews, the authors of [21] have used a logistic regression-based classifier. The reason for using logistic regression is that it estimates the probability whether a review is a spam review. Using a data set of 470 manually labeled reviews from Amazon product reviews, they have reported a very high accuracy (AUC=0.98) in separating non-reviews and brand reviews from legitimate reviews. They also experimented with support vector machine, decision tree, and naïve Bayesian classification. However, they have reported that such methods yielded poorer results.

Authors used a total of 36 features for supervised learning. These features have been divided into 3 categories: review-centric features, reviewer-centric features, and product-centric features. Some of those 36 features are mentioned below.

- **Review centric features (characteristics of reviews)**
  Review centric features include:
    1. the amount of feedback on a review
    2. the total amount of feedback that marked the review as helpful
    3. title length of the review
    4. the total length of the review
    5. the rating given by the review and its deviation from the average rating, etc.

  Review centric features also include some *textual features*, such as
    6. the number of positive and negative words, i.e., bad, beautiful, poor, etc.
    7. the cosine similarity between the review and product features. Product features were obtained from the product description text at amazon.com
    8. the percent of times the brand name has been mentioned in the review
    9. the percent of numerals
    10. the percent of capitals
    11. the percent of capital words

  According to authors, too many numerals or too many all-capital words signal a non-review.

- **Reviewer centric features (features related to Reviewers)**
  Reviewer centric features include
    1. the ratio between the number of reviews that a reviewer has written which are the first reviews of the products and the total number of reviews that the reviewer has written

2. the average rating of the reviewer
3. a feature indicating whether the reviewer always gave only good, average or bad rating, etc.

- **Product centric features (characteristics of products)**
  Product centric features include:
  1. the price of the product
  2. the sales rank of the product

The results of the experiment are given in table 6 below.

| Spam Type | AUC | AUC– textual features only | AUC – without feedback |
|---|---|---|---|
| Brand review and non-review | 98.7% | 90% | 98% |
| Brand review only | 98.5% | 88% | 98% |
| Non-review only | 99.0% | 92% | 98% |

Table 6: Results of Logistic Regression based Classifier for Detection of Spam Reviews [21]

Using only the textual features, the accuracy of classifier is not very good. The accuracy of classifier, when feedback on reviews is excluded, is still good. The reason for this is that feedback can be spammed too.

Finding untruthful reviews is difficult, even manually. The authors of [21] have used near-duplicate detection techniques to detect a subset of false reviews. They have also taken into consideration some suspicious cases, such as:

1. the same user-id posting the same review text to different products.
2. different user-ids posting the same review text on the same product.
3. different user-ids posting the same review text on different products.

The authors used 2-gram based review content comparison. Review pairs with similarity score of at least 90% were labeled as duplicates. In detecting spam reviews, they achieved the AUC of 0.78. The authors of [22] have also proposed similar methods for the detection of untruthful reviews.

*D. Voting Spam Detection*

The authors of [24] have proposed a voting method that analyzes the social network of users, and lowers the weight of votes received from the users that are not well connected to other users. They have shown results from preliminary experiments indicating that this method is effective in detecting and demoting content that is positively voted but is actually spam.

To detect voting spam, we propose a system in which more weight is given to the votes from users who are more credible. Credibility could be determined from a number of factors, such as

1. the total number of connections of the user.
2. the amount of content that the user has posted.
3. the number of positive votes versus the number of negative votes that the user has received on his content.
4. the amount of content posted by the user flagged as spam by other users.

*E. Hashtag Hijacking Detection*

The authors of [29] have proposed a machine learning based method which could be used to detect hashtag-based spam in twitter. The authors have observed in their twitter dataset, HSpam14, that spammers use more hashtags per tweet, probably for the purpose of reaching more users with a single spam tweet. They also observed that 90% of ham tweets (tweets that are not spam) used hashtags in lower case. Furthermore, they also observed that spam tweets often had an external link and the number of hashtags in many spam tweets was more than 1, etc.

After observing hashtag-based spam tweets carefully, they set forth a number of features which could be used to detect hashtag-based spam tweets. Some features were related to tweets, some were related to hashtag(s) being used in the tweet, and some were related to the spammer's profile on twitter. Among all those features, the authors selected top fifteen features for the detection of hashtag-based spam in tweets. Those features were selected on the basis of Gini coefficient. The higher the Gini coefficient of a feature, the more reliable it is for hashtag spam detection in tweets. The selected fifteen features are given in table 7 below.

| Rank | Feature of Tweet | Gini Coefficient |
|---|---|---|
| 1. | Presence of more than two hashtags in the tweet | 0.0405 |
| 2. | Presence of spammy hashtags in the tweet | 0.0175 |
| 3. | The tweet owner has less than 5 percentile followers | 0.0147 |
| 4. | Ratio between the number of Followers and Followees of the tweet owner | 0.0114 |
| 5. | Profile of the tweet owner has description | 0.0110 |
| 6. | Tweet owner has less than 5 percentile followees | 0.0096 |
| 7. | The tweet contains capitalized hashtags | 0.0062 |
| 8. | Fraction of capital letters in the tweet | 0.0060 |
| 9. | Presence of URL in the tweet | 0.0046 |
| 10. | Presence of exclamation mark in the tweet. | 0.0037 |
| 11. | Percentile of followers of the tweet owner | 0.0034 |
| 12. | The profile of the tweet owner has information about time zone | 0.0032 |
| 13. | Presence of negative sentiments in the tweet. | 0.0029 |
| 14. | The profile of the tweet owner contains URL | 0.0023 |
| 15. | Presence of suffix hashtag in the tweet | 0.0019 |

Table 7: List of Features Selected by Authors of [29] for hashtag-based spam Detection in Tweets.

To evaluate the effectiveness of these 15 features in detecting hashtag-based spam in tweets, the authors used a logistic regression classifier. For training the classifier, they used tweets posted on 17[th] May 2013. They manually labeled tweets as spam or ham (tweets that are not spam). For testing the classifier, they used tweets posted on 18[th] May 2013. According to authors, the classifier achieved precision of 0.96, recall of 0.77, and $F_1$ of 0.86.

*F. No Follow Attribute for Link Spam Prevention in Publicly Writable Pages*

Because of heavy link-based spamming in comments and in publicly writable pages, such as wikis, forums, etc., *no follow* attribute for hyperlinks has been introduced. If a hyperlink on a page has *no follow* attribute, it will be a signal for the search engine to not give any importance to this link while performing link-based ranking [12] [31]. An example of a hyperlink with *no follow* attribute is as follows:

```
<a href="target_page_link" rel="nofollow">
Click here to visit the source of information
</a>
```

Wikipedia, one of the most popular wikis, always uses *no follow* attribute in each reference. The use of *no follow* attribute by Wikipedia has been criticized by legitimate websites because *no follow* attribute prevents the deserved rank from being passed to the source of information (Source: Wikipedia).